\documentclass[manuscript]{aastex}
\usepackage{tikz}
\usepackage{amsmath}

\usepackage{graphicx}

\usepackage{pdflscape}
\usepackage[normalem]{ulem} 
\usepackage{soul} 
\usepackage{aas_macros}

\slugcomment{\textbf{ApJLetters} }

\shorttitle{3I/ATLAS}
\shortauthors{Jewitt}

\begin{document}

\title{Pre-perihelion Development of Interstellar Comet 3I/ATLAS}

\author{
David Jewitt$^{1}$ and Jane Luu$^{2}$
} 
\affil{$^1$Department of Earth, Planetary and Space Sciences, UCLA, 595 Charles Young Drive, Los Angeles, CA 90095\\}
\affil{$^2$ Centre for Planetary Habitability (PHAB), Department of Geosciences, University of Oslo, NO-0315 Oslo, Norway}

\email{djewitt@gmail.com}

\begin{abstract}
We describe pre-perihelion optical observations of interstellar comet 3I/ATLAS taken during July - September 2025  using the Nordic Optical Telescope.  Fixed aperture photometry of the comet is well described by a power law function of heliocentric distance, $r_H$, with the exponent (``index") $n = 3.8\pm 0.3 $ across the 4.6 au to 1.8 au  distance range (phase function 0.04$\pm$0.02 magnitude degree$^{-1}$ assumed). This indicates that the dust production rates vary in proportion to $r_H^{-1.8\pm 0.3}$.  An $r_H^{-2}$ variation is expected of a strongly volatile material, and consistent with independent spectroscopic observations showing that carbon dioxide is the primary driver of activity.  The measured heliocentric index is unremarkable in the context of solar system comets, for which $n$ is widely dispersed, and provides no basis on which to describe 3I as either dynamically old (thermally processed) or new (pristine). The morphology of the comet changes from a Sun-facing dust fan in the early 2025 July observations, to one dominated by an antisolar dust tail at later dates.  We attribute the delayed emergence of the tail to the large size (effective radius 100 $\mu$m) and slow ejection (5 m s$^{-1}$) of the optically dominant dust particles, and their consequently sluggish response to solar radiation pressure.   Small (micron-sized) particles may be present but not in numbers sufficient to dominate the scattering cross-section.  Their relative depletion possibly reflects interparticle cohesion, which binds small particles more effectively than large ones.
A similar preponderance of 100 $\mu$m grains was reported in 2I/Borisov.  However, 2I differed from 3I in having a much smaller (asteroid-like) heliocentric index, $n = 1.9 \pm 0.1$. Dust production rates in 3I are $\sim$180 kg s$^{-1}$ at 2 au, compared with $\sim$70 kg s$^{-1}$ in 2I/Borisov at the same distance. 
\end{abstract}


\section{INTRODUCTION}
\label{intro}
Comet C/2025 N1 (ATLAS) was discovered on UT 2025 July 1 as part of the ATLAS sky survey \citep{Den25}. Pre-discovery observations soon confirmed the orbit to be strongly hyperbolic, leading to its recognition as the third known interstellar interloper, after 1I/'Oumuamua and 2I/Borisov.  With eccentricity $e$ = 6.145, inclination $i$ = 175.1\degr~and perihelion at $q$ = 1.357 au (solution reference JPL\#13, data arc from 2025 May 21 to 2025 July 18), the orbit of the renamed object 3I/ATLAS (hereafter simply ``3I''), is by far the most extreme of any yet recorded in the solar system, with a velocity at infinity of approximately 58 km s$^{-1}$.  The perihelion date is UT 2025 October 29.

It is statistically unlikely that 3I has approached any star more closely than it will approach the Sun in 2025 (see Section 3.1), at least since the (presumed) ejection from its parent protoplanetary disk.  Therefore, pre-perihelion observations present an opportunity to study the rise of activity on an object that has been held at interstellar temperatures ($\lesssim$10 K) possibly for billions of years \citep{Tay25}.  Here we present systematic observations taken to assess the development of activity in 3I on the way to perihelion and to determine basic dust parameters from its changing morphology.  We compare 3I with published observations of the second interstellar interloper, 2I/Borisov, and with lightcurve observations of solar system comets analysed by \cite{Lac25}.

\section{OBSERVATIONS}

We observed 3I using the 2.5m diameter Nordic Optical Telescope (NOT) located at 2400m altitude in La Palma, Canary Islands, Spain.  The observations were taken in target-of-opportunity mode and used charge-coupled device (CCD) detectors already mounted on the telescope for other science purposes.  Our early observations were taken with STANCAM, while later ones used ALFOSC.

1) The STANCAM is a Tektronix 1024x1024 pixel back-side illuminated and thinned CCD with 24 $\mu$m pixels each subtending 0\arcsec.176 in the focal plane.  
The field of view is approximately 3\arcmin$\times$3\arcmin.  

2) The Alhambra Faint Object Spectrograph and Camera (ALFOSC) uses an e2v Technologies 2048 x 2064 pixel back-illuminated CCD, with each 15 $\mu$m pixel subtending 0\arcsec.214. The ALFOSC field of view is approximately 7\arcmin$\times$7\arcmin.  

On each night we obtained flat field exposures from an illuminated patch inside the dome, and a set of bias frames.  Our photometric monitoring observations  used the broadband Bessel R$_{641-148}$ filter, which has a transmission peak near 5900\AA~and half-power transmission at 5680\AA~and 7180\AA.  The use of the R filter minimizes the risk of contamination of the photometry by gas, because strong cometary resonance fluorescence lines from CN, C$_2$ and C$_3$ are  confined to wavelengths $<$5000\AA. 

Photometric calibration was obtained from measurements of nearby Landolt stars \citep{Lan92} having roughly Sun-like colors, supplemented by measurements of field stars from the Pan STARRS and GAIA sky catalogs.  A journal showing the  geometry for each successful observation is given in Table \ref{geometry}.

\subsection{Photometry}
The telescope was tracked non-sidereally to follow 3I, while guiding on field stars.  For each visit, we obtained consecutive images of 3I with integration times from 30s to 60s.   The short integration times were chosen to minimize trailing between the object and field stars.  In observations from early July, the accuracy of the photometry was limited to $\pm$0.05 to $\pm$0.1 magnitudes by the high density of field stars, as a result of the low (near 0\degr) Galactic latitude.  The Galactic latitude increased to 16\degr~by August 1, resulting in a concomitant decrease in the star density, a reduction of confusion with background sources and photometric scatter reduced to $\pm$0.01 to 0.02 magnitudes.  Where possible, we digitally removed field stars falling in the photometry aperture.  Measurements from individual images on a given night were generally consistent, and also consistent with photometry obtained from compiled nightly median images.  In a few severe cases of confusion with field stars, we conservatively chose to discard the data rather than attempting heroic correction efforts.  

Compared to field stars, 3I appears extended in all images as a result of the presence of near-nucleus dust.  We elected to use a fixed linear (as opposed to angular) photometry aperture in order to avoid the ``delta effect'', in which the photometry is affected by the particular radial surface brightness distribution in the coma \citep{Mar80}.  By trial and error, we chose a projected linear radius of 10$^4$ km, corresponding to $\sim$4\arcsec~on July 2 and $\sim$5.5\arcsec~on September 30; substantially smaller apertures are affected by variations in the seeing from night to night (mostly in the range from 1\arcsec~to 2\arcsec~FWHM (Full Width at Half Maximum)), while substantially larger apertures suffer disproportionately from photometric contamination by field stars. The sky background was determined from the median signal in a concentric annulus with projected inner and outer radii typically 20,000 km to 30,000 km.  Use of the median sky signal removes contamination by field stars and cosmic ray strikes.  We checked  the accuracy of the sky annulus measurement by taking spot measurements around 3I. 

\begin{figure}
\epsscale{0.99}

\plotone{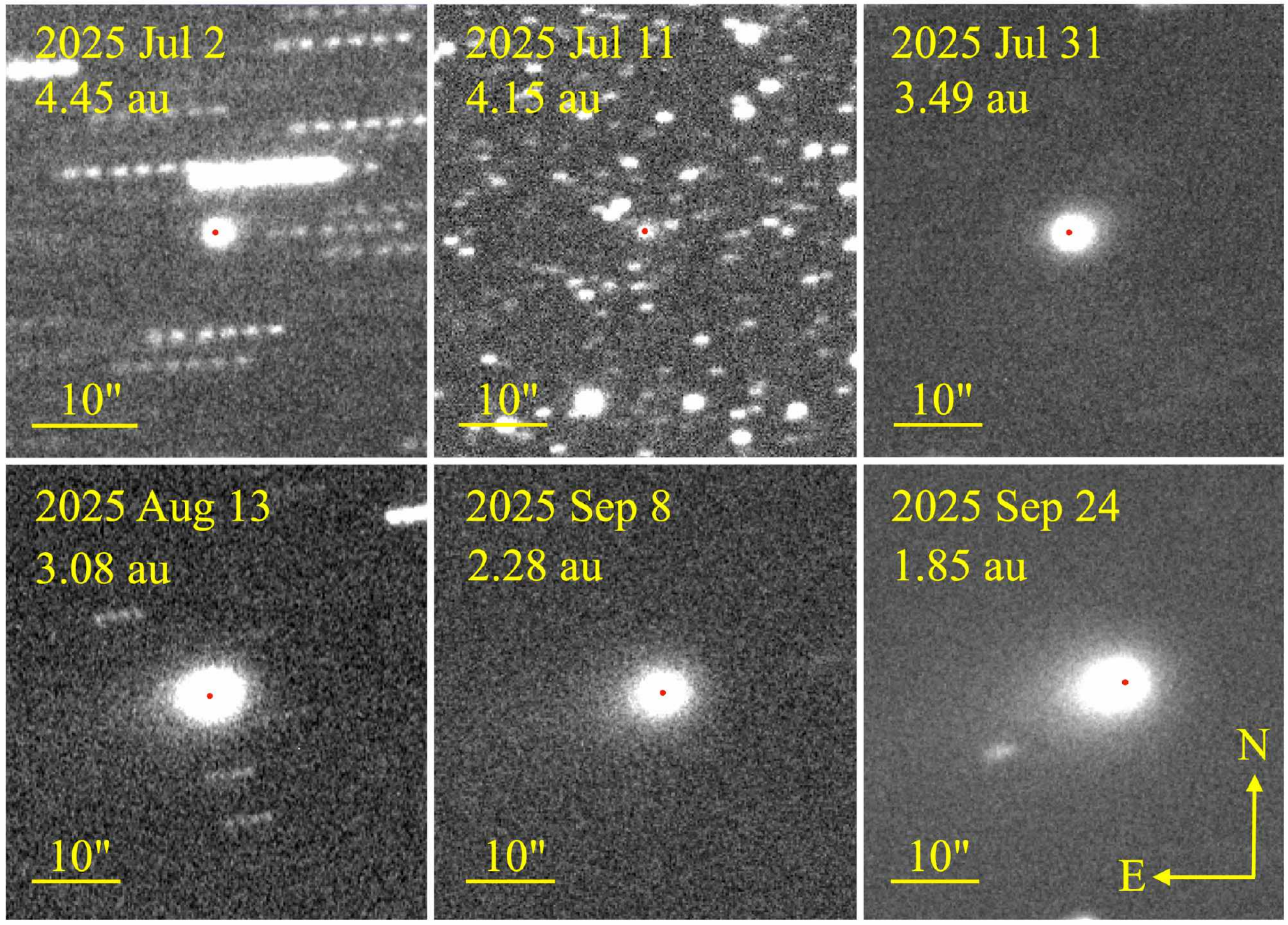}

\caption{Sample images showing the morphological development of 3I/ATLAS.  The image dates and heliocentric distances of the comet are shown in each panel, as is a 10\arcsec~scale bar.  Sunward is West.  The location of the brightest pixel is marked with a red dot, showing that the dust is extended asymmetrically towards the West in the early data, reversing to the East as the radiation pressure swept tail develops in the later data.   See Table \ref{geometry} for additional details.  \label{composite}}
\end{figure}

A composite of images is shown in Figure \ref{composite}, together with 10\arcsec~scale bars. All panels are averages of separate integrations (c.f., column 5 of Table \ref{geometry}) except the panel for 2025 July 11, which shows the single image obtained on that night without star overlap.  On each date the brightest pixel is marked with a red dot, showing that the dust is distributed more to the west in the July data, with no evident tail, then shifting to the east as a radiation pressure swept tail develops in August and September (see Section \ref{dusty}).  The photometry is summarized in Table \ref{photometry}.  Errors on the apparent magnitude, $m_R$, were obtained from repeated observations on a given night and do not include possible systematic errors.  Statistical errors on the absolute magnitudes, $M_R$, are the same as on $m_R$, but additional systematic errors exist owing to the unmeasured phase function.  

\clearpage

\section{DISCUSSION}
\subsection{Heliocentric Lightcurve}
The nucleus of 3I contributes negligibly to the photometry \citep{Jew25}. Instead, the apparent magnitude, $m_R$, measures the dust content of the coma and is affected by the geometry of the observation.   We write 

\begin{equation}
m_R = M_R + 2.5 \log_{10}(r_H^n \Delta^m) - 2.5 \log_{10}(\Phi(\alpha))
\label{invsq}
\end{equation}

\noindent where $r_H$ and $\Delta$ are the heliocentric and geocentric distances in au and $\alpha$ is the phase (Sun-object-Earth) angle.  Heliocentric magnitude, $M_R$, and indices $n$ and $m$ are properties of the target.  For an asteroidal (fixed cross-section) body, the heliocentric and geocentric indices, $n$ and $m$, respectively, are both $n = m = 2$, and Equation \ref{invsq} reduces to the familiar inverse square law of brightness variation, albeit modified by the angle-dependent phase function, $\Phi(\alpha)$.  

For a resolved object measured with an aperture of fixed angular radius, the index $m$ depends on the  distribution of the surface brightness, because larger apertures measure scattering from particles within a larger volume around the nucleus as $\Delta$ increases.  For example, a comet outgassing in steady state produces a coma surface brightness inversely proportional to the angular distance from the nucleus and, for such a distribution, the resulting geocentric index is $m$ = 1 \citep{Jew87}.   Since $r_H$ and $\Delta$ are partially correlated, the choice of $m$ can affect the derived value of $n$ in fixed angular radius photometry (smaller $m$ producing larger $n$). To avoid this problem,  we here use photometry apertures of 10$^4$ km fixed \textit{linear} (as opposed to angular) radius when projected to the comet.  This yields a measure of the scattering cross-section within a fixed volume around the nucleus, for which no delta effect exists and $m$ = 2 can be safely assumed in Equation \ref{invsq}.   

The phase function, $0 \le \Phi(\alpha) \le 1$, measures the angle dependent variation of the scattering efficiency of cometary dust, normalized at phase angle $\alpha$ = 0\degr.  This phase function is unmeasured in 3I.  Based on experience with other comets and consistent with the recent summary by \cite{Ber25}, we approximate -2.5$\log_{10}(\Phi(\alpha)) = \beta_R \alpha$, with nominal $\beta = 0.04$, over the limited phase angle range of the 3I data (c.f., Table \ref{geometry}).  We assume representative values of the coefficient $\beta_R$ = 0.02, 0.04 and 0.06 magnitudes degree$^{-1}$ to examine the effect of the phase function on the interpretation of the data.

With these assumptions, we re-write Equation \ref{invsq} as


\begin{equation}
M_R = m_R - 2.5n\log_{10}(r_H) - 5\log_{10}(\Delta) - \beta_R \alpha
\label{invsq2}
\end{equation}

\begin{figure}
\epsscale{0.9}

\plotone{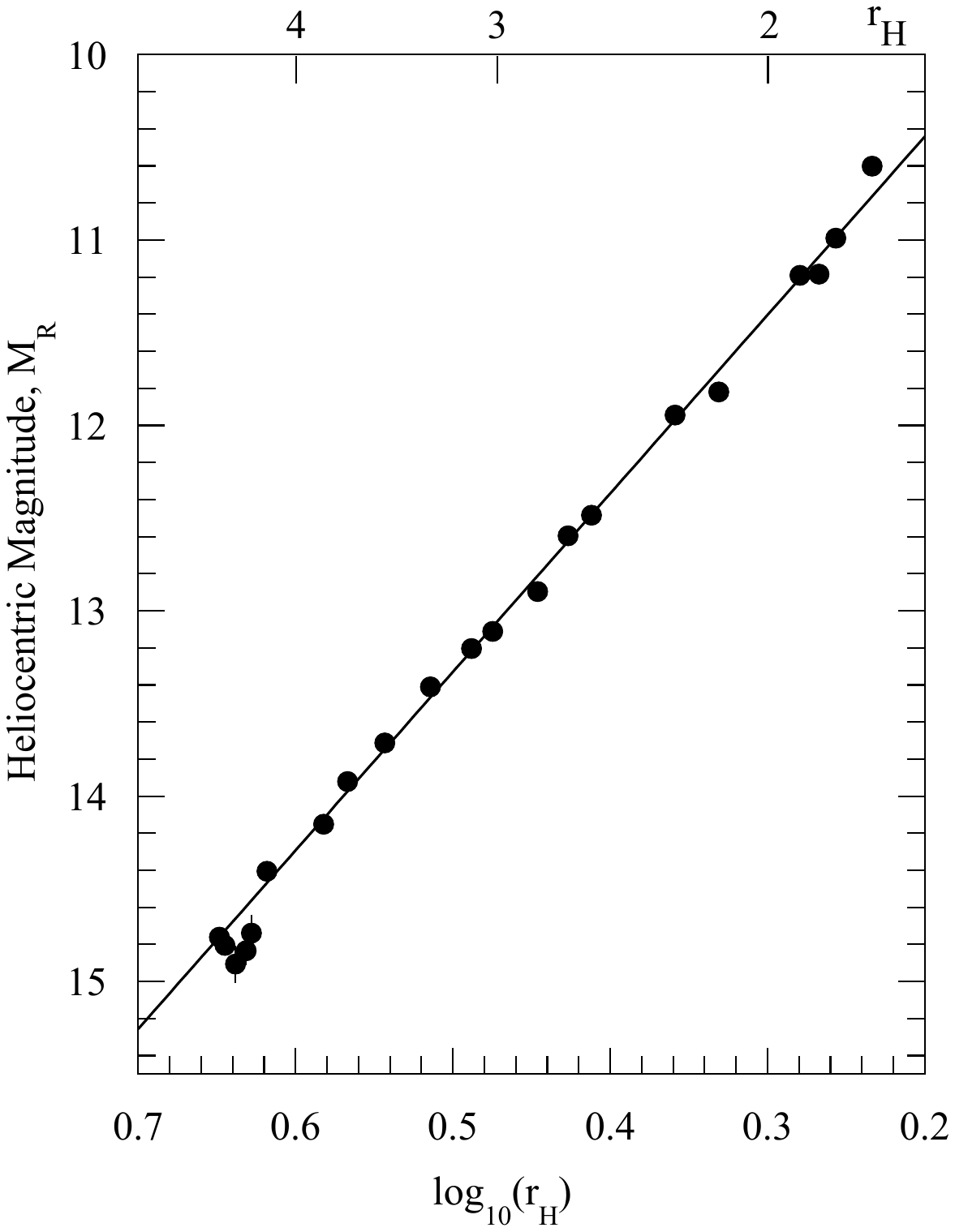}

\caption{Reduced photometry from NOT, assuming $\beta_R$ = 0.04 magnitudes degree$^{-1}$, as a function of heliocentric distance.  The straight line is a least squares fit of slope $n$ = 3.8.  \label{Photometry_Plot}}
\end{figure}

\noindent We plot $M_R$ as a function of $\log_{10}(r_H)$ in Figure \ref{Photometry_Plot}, where the solid line  is a weighted least-squares fit to the data.  The fit with $\beta_R$ = 0.04 magnitudes degree$^{-1}$ gives $M_R(r_H=1)$ = 8.59$\pm$0.02 and $n$ = 3.79$\pm$0.02. 
These tiny formal errors reflect only uncertainties on measurements of $m_R$ and therefore underestimate the true uncertainties introduced by the adoption of an unmeasured phase function.  To estimate more realistic uncertainties, we recomputed the fits using $\beta_R$ = 0.02 magnitudes degree$^{-1}$ and 0.06 magnitudes degree$^{-1}$ and used the differences in the fit parameters to estimate the uncertainties.  By this procedure we find that the NOT data are broadly represented by $M_R(r_H=1)$ = 8.6$\pm$0.7 and $n$ = 3.8$\pm$0.3, across the pre-perihelion 4.6 au to 1.8 au range.

Other observatories have reported 3I lightcurve data, albeit over different ranges of heliocentric distance, in different filters, with different aperture sizes and shapes and measured in different ways (notably using fixed angle vs.~linear radius apertures).  We briefly describe these other reported photometric datasets.

\textbf{Zwicky Transient Facility (ZTF):} ZTF uses a 1.5 m aperture telescope with 1\arcsec~pixels and filters approximating g and r of the Sloan filters \citep{Yor00}.  Prediscovery  detections  of 3I were obtained starting UT 2025 May 15 \citep{Ye25}. The ZTF photometry used an aperture of fixed angular radius 3\arcsec~and the reported V-band magnitudes were ``corrected'' for variable seeing  assuming a model of the surface brightness of the coma. We digitized the data from Figure 3 of \cite{Ye25}.

\textbf{Asteroid Terrestrial-impact Last Alert System  (ATLAS):}  ATLAS is a distributed network of 0.5 m diameter telescopes with 1.9\arcsec~pixels and three broadband optical filters.  The earliest 3$\sigma$ detection of 3I was made UT 2025 April 27. We used ``c'' (roughly 4200\AA~to 6500\AA) and ``o'' (5600\AA~to 8200\AA) filter ATLAS photometry from the ``m10'' (fixed 10\arcsec~wide, square photometry aperture) column in Table 1 of \cite{Ton25}.  

\textbf{Transiting Exoplanet Survey Satellite (TESS):}  TESS uses 0.1 m diameter orbiting telescopes having 21\arcsec~pixels and filter bandpass from approximately 6000\AA \ to 10000\AA.  TESS measurements of 3I were reported by \cite{Fei25} and \cite{Mar25}.  We use photometry from Table 2 of the latter paper obtained with a fixed angular 63\arcsec~radius aperture.

\begin{figure}
\epsscale{0.9}

\plotone{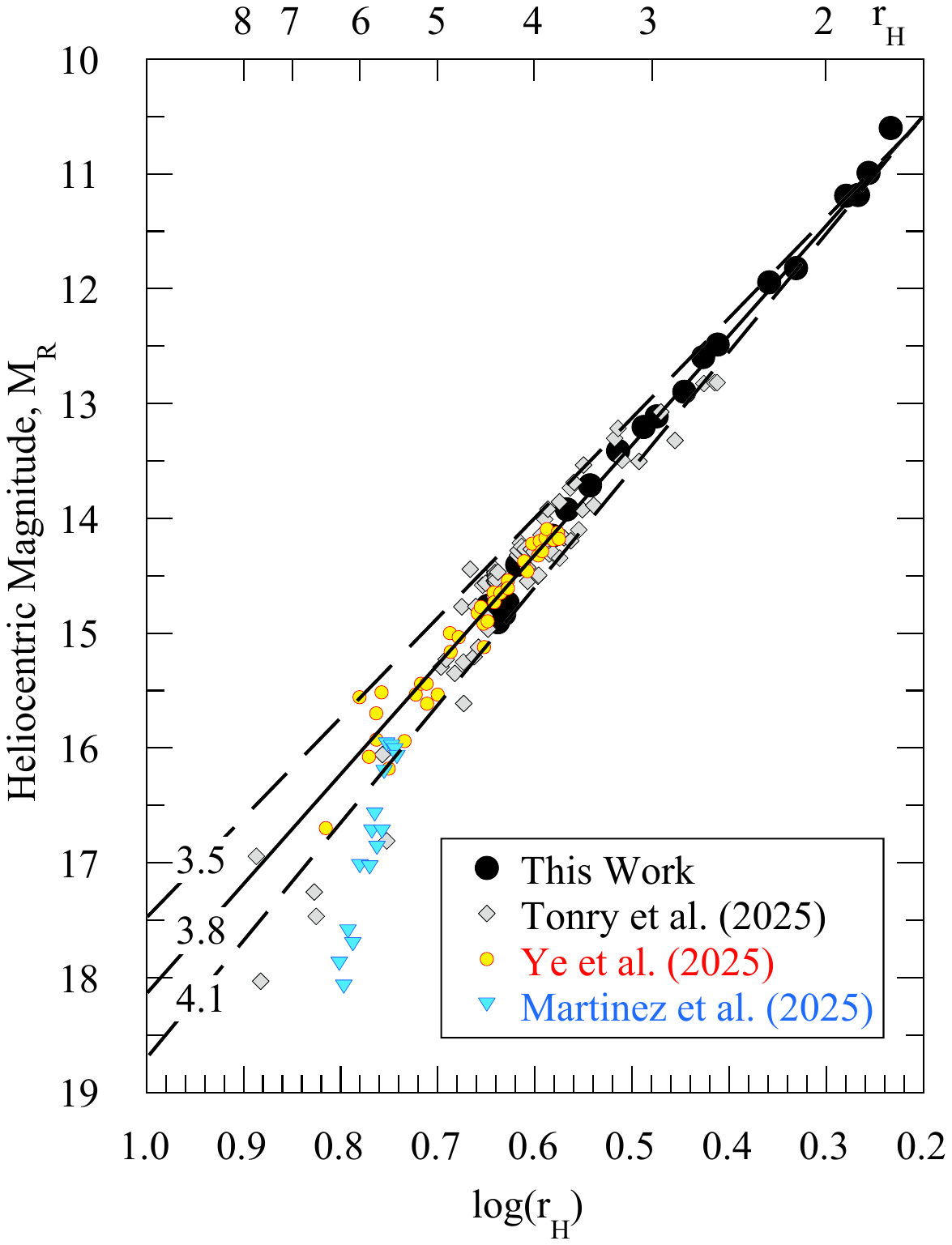}

\caption{Composite pre-perihelion lightcurve of 3I/ATLAS with error bars omitted for clarity.  The solid line shows heliocentric index $n$ = 3.8 while the dashed lines above and below it represent the $\pm1\sigma$ uncertainties on $n$.  Error bars are omitted for clarity of presentation. \label{Summary_plot}}
\end{figure}

We compare the different datasets in Figure \ref{Summary_plot}. 
The different  apertures, filters and measurement schemes lead to systematic offsets between the lightcurves which we have removed by shifting the reported magnitudes to align with the NOT data.  Specifically, the data from ATLAS \citep{Ton25} were shifted brighter by 0.04 magnitudes and those from ZTF \citep{Ye25} by 0.68 magnitudes, while data from TESS \citep{Mar25} were shifted fainter by 0.42 magnitudes in order to smoothly match the NOT lightcurve.  Results of least squares fits to the photometry are listed in Table \ref{fits}.  In each case, we weighted the data by the formal measurement uncertainties; we ignored reported upper limits to the brightness, and included uncertainties resulting from the adoption of assumed phase coefficients $\beta_R$ = 0.04$\pm$0.02 magnitudes degree$^{-1}$.   Overall, we find good agreement between the heliocentric indices derived from the NOT, ATLAS and ZTF photometry for $r_H \lesssim$ 5 au and we conclude that the heliocentric variation, $n = 3.8\pm$0.3, is well determined in this distance range.  However, the ATLAS, ZTF and TESS datasets are not consistent with each other at $r_H \geq 5.6$ au (Figure \ref{Summary_plot}) and we consider this portion of the lightcurve indeterminate.  

While index $n$ = 2 corresponds to a constant scattering cross-section in the photometry aperture, $n$ = 3.8 implies a cross-section that varies as $r_H^{-1.8}$.  This is a less steep dependence than expected for water ice, whose rate of sublimation varies as $r_H^{-3.4}$ to $r_H^{-14.5}$ across the 1.8 $\le r_H \le$ 4.4 au range (depending on the distribution of absorbed solar power; c.f., Figure 1 of \cite{Jew25c}). However, sublimation of carbon dioxide is already known as the dominant driver of mass loss from 3I at $r_H$ = 3.32 au  (\citet{Lis25}, \citet{Cor25}), and CO$_2$ sublimates in proportion to $r_H^{-2}$ in this distance range (c.f., \cite{Jew25c}) providing a simple explanation of the measured index.  We do not see evidence for a break in the heliocentric index that would be expected if, for example, the dominant sublimating volatile changed from CO$_2$ to H$_2$O in the 4.5 au to 1.7 au heliocentric distance range covered by the NOT data.  Hopefully, enough spectroscopic measurements have been taken in this range to show whether or not this change happened.

Hubble Space Telescope observations set an upper limit to the radius of the 3I nucleus at $r_n <$ 2.8 km \citep{Jew25}, corresponding to magnitude at $r_H$ = $\Delta$ = 1 au and phase $\alpha$ = 0\degr~of $M_R \sim$ 15.8. (Note that \cite{Clo25} reported a consistent but very similar \textit{lower} limit to the radius $r_n > 2.5$ km using a non-detection of nongravitational acceleration).  The nominal fit parameters from the NOT photometry (Table \ref{fits}) reach $M_R = 15.8$ at $r_H$ = 5.7 au, while if the radius were instead $r_n$ = 1 km (corresponding to $M_R \sim$18.0) the distance would be $r_H \sim$ 9.8 au.  These (very approximate) estimates of the activity initiation distance are compatible with the (equally approximate) $\sim$6 au and $\sim$9 au turn-on distance estimates by \cite{Fei25} and \cite{Ye25}, respectively, and with early activity driven by CO$_2$ but not H$_2$O sublimation.  The steeply rising TESS photometry (Figure \ref{Summary_plot}) suggests a smaller distance, $r_H \sim$ 6.3 au, but the discordance with other datasets beyond $r_H \geq 5.6$ au is again noted.

\begin{figure}
\epsscale{1.0}

\plotone{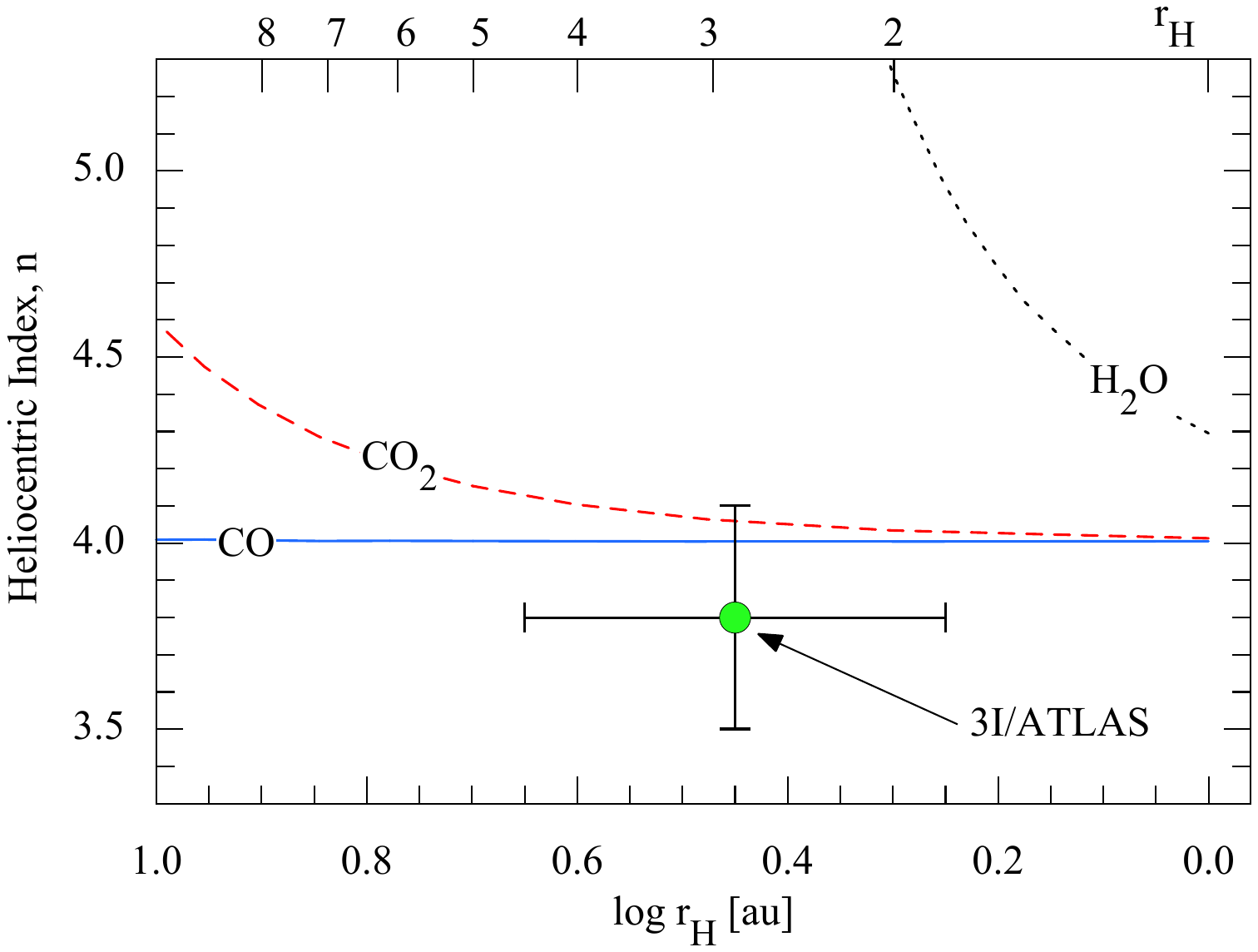}

\caption{Heliocentric indices computed from equilibrium sublimation models for CO, CO$_2$ and H$_2$O ices as a function of heliocentric distance.  Hemispherical sublimation is assumed, and the ices are taken to be perfectly absorbing and emissive.  The measured index for 3I/ATLAS is shown.  The vertical error on the 3I point denotes the $\pm$0.3 uncertainty on the heliocentric index and shows consistency with free sublimation of CO and CO$_2$ but not H$_2$O.  The horizontal bar marks the range of distances over which 3I was reliably observed (c.f., Table \ref{geometry}).   \label{indices}}
\end{figure}

We solved the energy balance equation for exposed, sublimating ice (as described in section 3.4 of \cite{Jew19}), using the physical constants for CO, CO$_2$ and H$_2$O (\cite{Bro80}, \cite{Was26}). For each ice, we calculated the specific sublimation mass flux, $f_s$ [kg m$^{-2}$ s$^{-1}$] as a function of heliocentric distance, and then calculated the derivative, $df_s/dr_H$, to compare with the heliocentric index of 3I.  For simplicity, we assumed that the ice is perfectly absorbing (Bond albedo  = 0) and emissive (emissivity = 1) and we present models for hemispherical (dayside) heating, motivated by the sunward launch of dust into the sunward fan.  The results are shown in Figure \ref{indices}, where it is evident that sublimating  CO and CO$_2$ are both consistent with the observed heliocentric variation in 3I. H$_2$O sublimation is  confidently ruled out as an explanation of the pre-perihelion lightcurve of 3I.  It is clear that the 2D equilibrium sublimation model is overly simplistic. A more  realistic 3D model for CO and CO$_2$ should include heat conduction into, and gas escape from, a porous thermal skin on the nucleus, allowing  volatile species to be liberated at lower temperatures from beneath the physical surface.  Such a model would depend on unmeasured values of both the thermal diffusivity and porosity of the nucleus, and so would itself be under-constrained.  

\begin{figure}
\epsscale{0.9}

\plotone{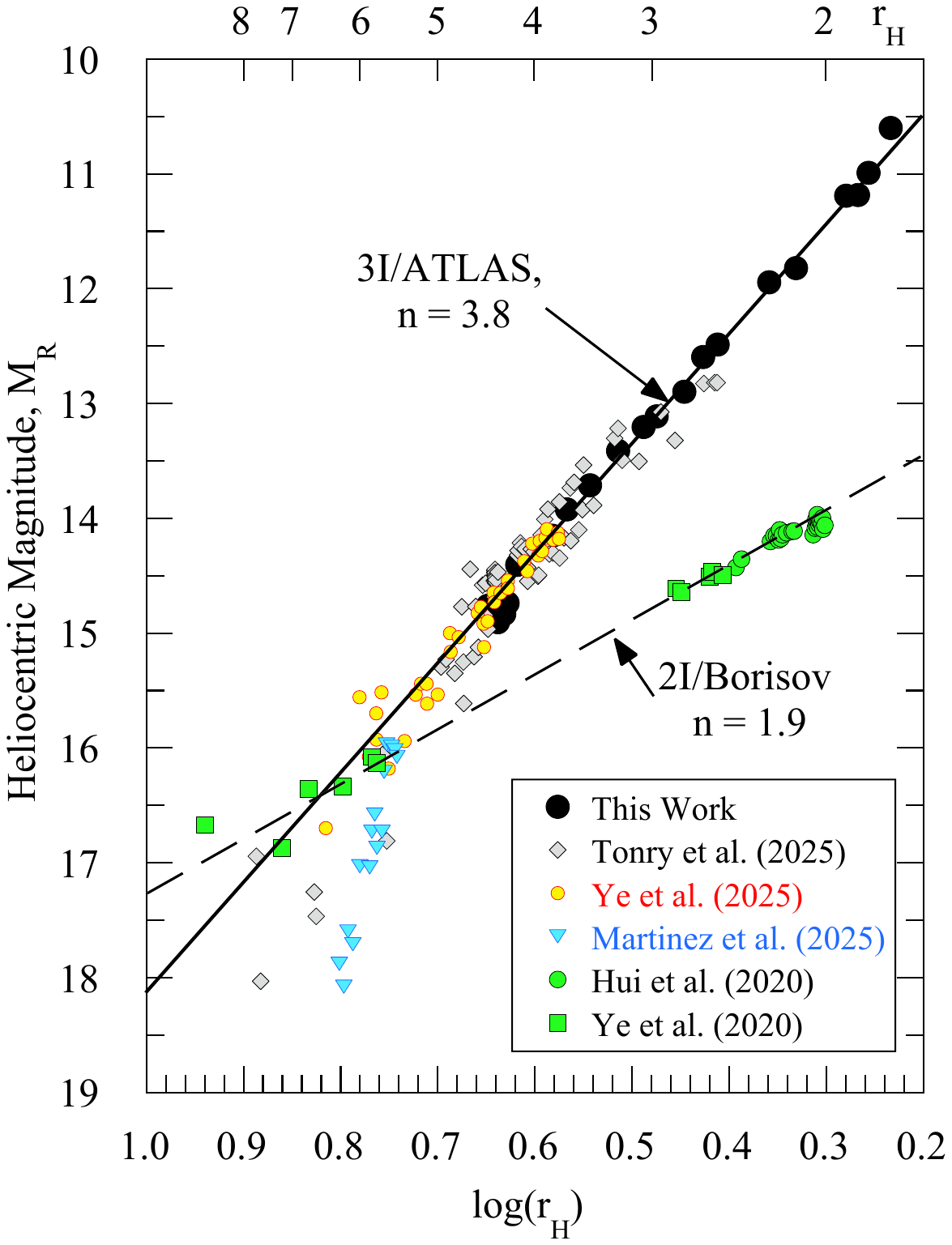}

\caption{Composite pre-perihelion lightcurves of 2I/Borisov and 3I/ATLAS compared, with error bars omitted for clarity.  The best-fit heliocentric indices are marked for each object.  \label{Summary_Borisov}}
\end{figure}

We also show for comparison in Figure \ref{Summary_Borisov}  the reported pre-perihelion lightcurve of the second interstellar interloper, 2I/Borisov.  Measurements from \cite{Hui20} were taken with a fixed projected  aperture radius of 10$^4$ km, while those from ZTF \citep{Ye20} employed a fixed angular aperture of 5\arcsec~radius, projecting to 21,000 - 28,000 km at 2I.  We shifted the ZTF data fainter by 0.24 magnitudes to match the \cite{Hui20} photometry and, with this shift, the two datasets are in good agreement.  Separate fits to the two datasets give a mean $n $ = 1.9$\pm$0.1, a much smaller index than for 3I (see Table \ref{fits}).  This small index is consistent with a cross-section independent of $r_H$, which is difficult to reconcile with the expected rise of activity towards perihelion.  One speculative possibility is that the dust grains in 2I/Borisov were themselves volatile.  If so, grains of a given size would  sublimate on a timescale varying as $r_H^{2}$.  Then, with the production rate increasing towards perihelion as $r_H^{-2}$, and their lifetime against sublimation decreasing as $r_H^2$, the instantaneous mass and cross-section of the coma would be approximately independent of heliocentric distance. Opposing this idea is the observation that water ice absorptions were not detected in 2I/Borisov \citep{Yan20}.

\begin{figure}
\epsscale{0.99}

\plotone{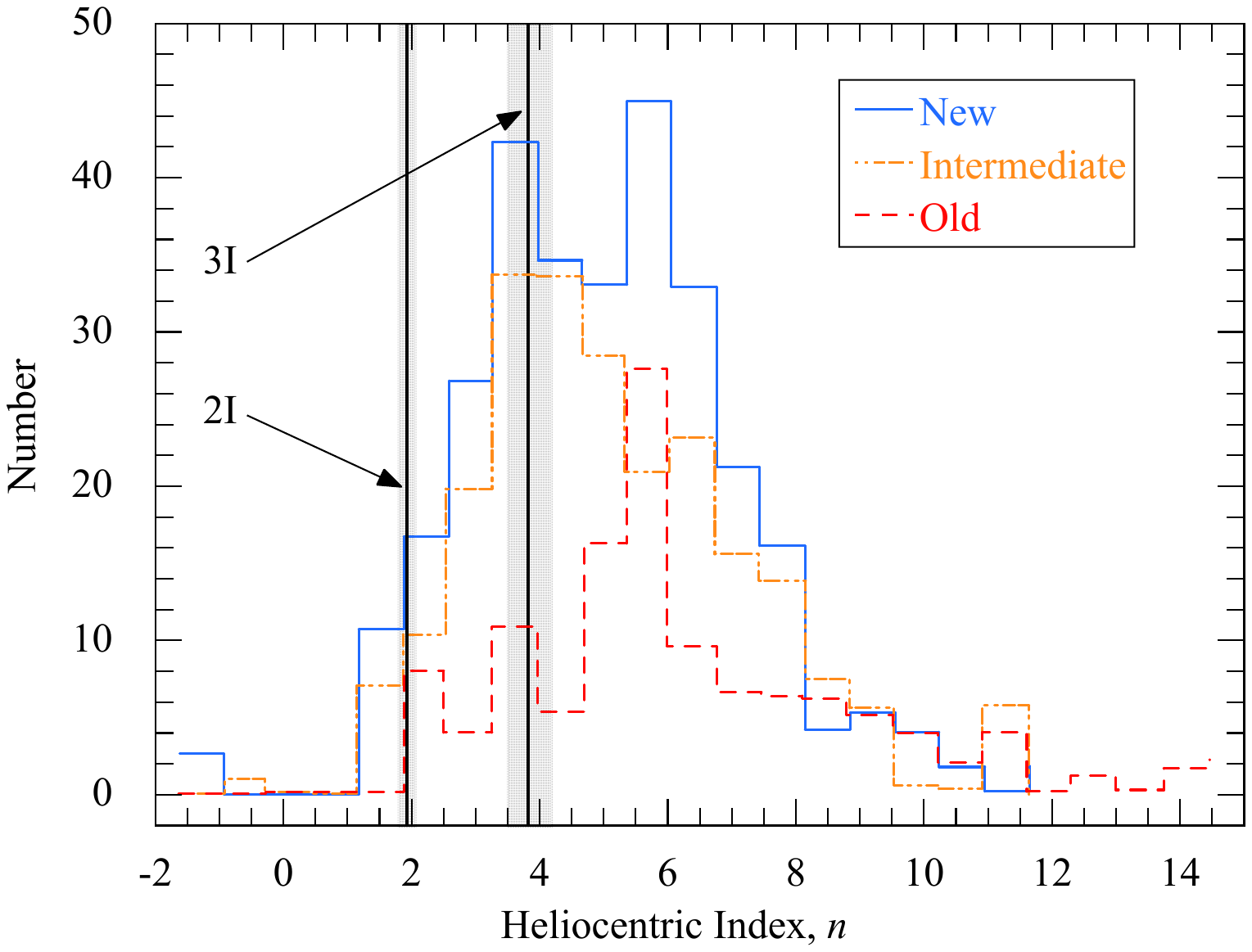}

\caption{The heliocentric brightness indices of 2I and 3I, $n$ = 1.9$\pm$0.1 and 3.8$\pm$0.3, respectively, compared with histograms of the same quantity for comets in three dynamical groups.  Dynamically new (solid blue line), intermediate (orange dash-dot line) and old (red dashed line) are shown.  The formal uncertainties on the indices for 2I and 3I are shown as shaded regions.  Comet data are from \cite{Lac25}. \label{Lacerda_plot}}
\end{figure}

\cite{Ye25} reported that the heliocentric index of 3I is more similar to the indices of short-period and dynamically old long-period comets than to dynamically new long-period comets from the Oort cloud.
These authors suggested that the index is a measure of past thermal processing, and that 3I might have been heated by previous stellar flybys\footnote{ Statistically, individual interstellar interlopers are extremely unlikely to have previously approached any star. To see this, note that the timescale for passing within distance $d$ of a star is given by $\tau \sim (N_1 \pi d^2 \Delta V)^{-1}$, where $N_1$ is the number density of stars and $\Delta V$ is the  velocity of the interstellar object relative to the stars.  If we represent the Milky Way as a collection of $N_{\star}$ stars, distributed uniformly in  a disk with radius, $R_G$, and thickness, $H_G$,  then $N_1 = N_{\star}/(\pi R_G^2 H_G)$ and $\tau \sim R_G^2 H_G/(N_{\star} d^2 \Delta V)$. We set $N_{\star} \sim 10^{11}$, $R_G \sim$ 10 kpc,  $H_G \sim$ 1 kpc,  $\Delta V$ = 60 km s$^{-1}$ and  $d$ = 30 au, the radius of Neptune's orbit.  These values give $\tau \sim 2\times10^{19}$ s, or about 50 times the age of the universe. The timescale to approach a star within 1 au would be $\sim10^3$ times longer.} or when in the protoplanetary disk where it formed. Figure \ref{Lacerda_plot} shows the distribution of pre-perihelion heliocentric indices for dynamically new, intermediate and old comets, taken from \cite{Lac25}.  We used so-called ``strand'' data from their work to provide the most robust estimate of the pre-perihelion brightening rates in each dynamical group.   The Figure shows that the heliocentric index of 3I is unremarkable when compared to the distributions of the index of any of the three dynamical groups.  According to Figure 5 of \cite{Lac25}, some 36\% 
of dynamically new comets have $n \le$ 3.8 while this fraction for dynamically old comets is 21\%.  Thus, we find no basis for assigning a dynamical age, or any metric of past heating, to 3I based on its heliocentric index.  

The smaller heliocentric index of 2I/Borisov is more unusual.  Only 9\% of dynamically new and 5\% of dynamically old comets have $n \le$ 2.0.  Again, while  heliocentric indices as small as that in 2I/Borisov are rare in measured solar system comets, the evidence cannot be reliably used to infer a dynamical age classification.

\subsection{Dust Properties}
\label{dusty}
\begin{figure}
\epsscale{0.85}

\plotone{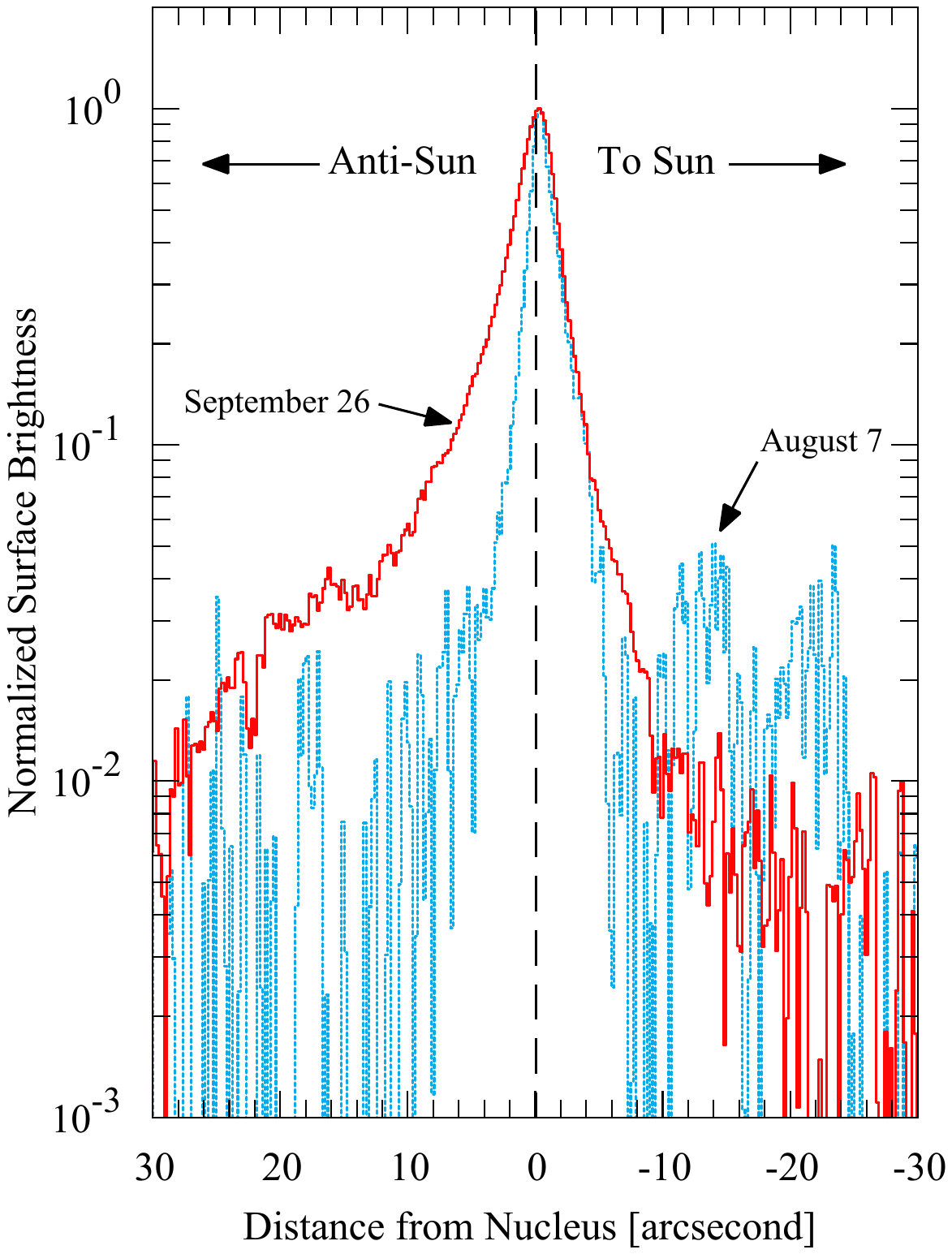}

\caption{The normalized surface brightness as a function of angular distance (positive towards the East) from the nucleus measured in a 2.6\arcsec~wide strip along the projected antisolar direction.  On August 7 (blue line) the profile is asymmetric with an extension towards the Sun, best seen in the inner few arcseconds.  By September 26 (red line) the sense of the asymmetry has reversed, and the antisolar tail is dominant.  \label{2SB_plot}}
\end{figure}

The earliest NOT observations of comet 3I (UT 2025 July 2, DOY$_{25}$ = 183 showed, upon close examination, a faint coma in the west (sunward) direction but no anti-sunward tail (\cite{Jew25b}, see Figure \ref{composite}).  The same sunward extension was reported almost simultaneously by \cite{Sel25} and \cite{Bol25}. Syndynes and synchrones computed for these early dates all project to the east, meaning that the sunward material cannot be an effect of projection, instead indicating a real projection of material towards the Sun \citep{Jew25}.  Sunward ejection from comets is completely normal, where it results from the preferential sublimation of ices on the hot, Sun-facing day side of the nucleus (e.g., \cite{Sek87}).  What is unusual in 3I is the relative weakness of any anti-sunward tail in the early imaging observations.  However, by mid-August (when $r_H \sim$ 3.0 to 3.5 au), the morphology had evolved to present a dominant tail of particles roughly aligned with the eastward (antisolar) direction, and this tail brightened and lengthened towards the last pre-perihelion observations in September (Figure \ref{2SB_plot}).  

\begin{figure}
\epsscale{0.8}

\plotone{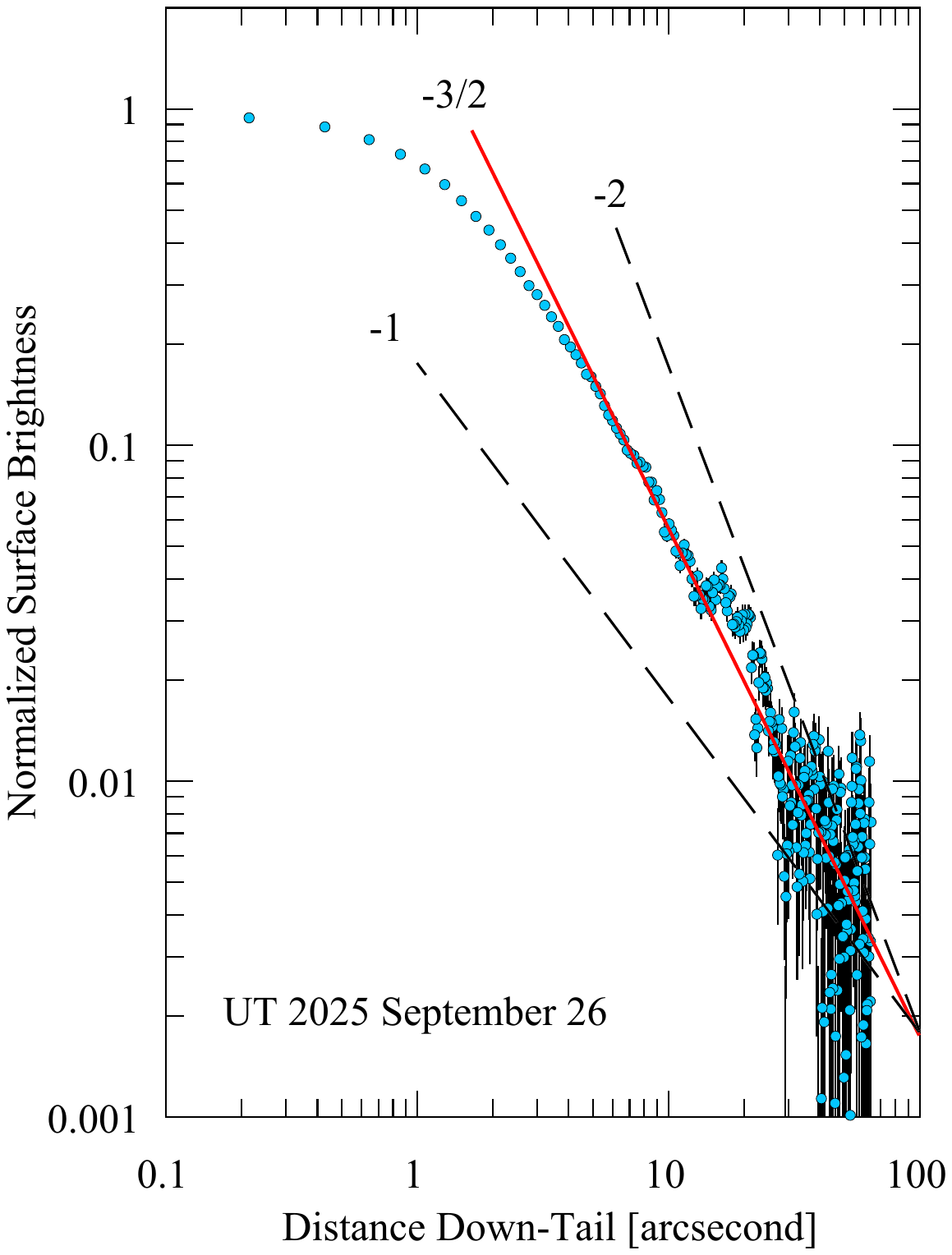}

\caption{Normalized surface brightness profile of the antisolar tail on UT 2025 September 26, with error bars computed from the standard deviation of nearby sky brightness measurements.   Lines of logarithmic slope -1, -3/2 and -2 are marked for reference. \label{downtail}}
\end{figure}

As a check of the radiation pressure origin of this emergent tail, we measured the surface brightness profile  using data from UT September 26.  First, we combined three 30s integrations on the comet and rotated the image to bring the tail axis to the horizontal.  Then we sampled the surrounding sky and its uncertainty using the mean and standard deviation of the signal within 27 star-free positions, each of 5$\times$5 pixels (1.1\arcsec$\times$1.1\arcsec).  After sky subtraction, we extracted the average signal from a 12-pixel (2.5\arcsec) wide strip along the tail axis.  This is shown in Figure \ref{downtail} normalized at the peak.  The profile at distances $\lesssim$ 5\arcsec~is affected by convolution with the wings of the seeing disk (FWHM = 2.1\arcsec~on this night, in observations taken at airmass $\sim$4.7), but at larger distances the profile converges on a power law  of slope $-1.5$, which is the value expected from the action of radiation pressure on a dust particle tail \citep{Jew87}. A weighted fit using the formal errors determined from sky measurements has slope -1.40$\pm$0.02.  However, the data are clearly non-Gaussian (e.g., see the bump near 16\arcsec~in Figure \ref{downtail} caused by an imperfectly removed background object) and the actual uncertainty on the slope must be larger than the formal $\pm$0.02.  We conclude that the profile is consistent with the action of radiation pressure on dust from 3I.  Our NOT data provide no evidence for sublimating grains in the tail, whose signature would be a progressive steepening of the gradient with increasing separation from the nucleus.   We infer that, if initially present, volatile rich grains  leaving the nucleus must have lost those volatiles before reaching the tail.  

To measure the position angles of the asymmetries we first formed nightly composite images and convolved these with a Gaussian function having a FWHM of 5 pixels (1.07\arcsec) to reduce noise.  The position angles and their uncertainties were measured relative to the center of light in each composite. Some nightly composites were not usable for this purpose owing to confusion with nearby field stars or internally scattered light, or simply from bright sky associated with scattered moonlight.    

\begin{figure}
\epsscale{0.8}

\plotone{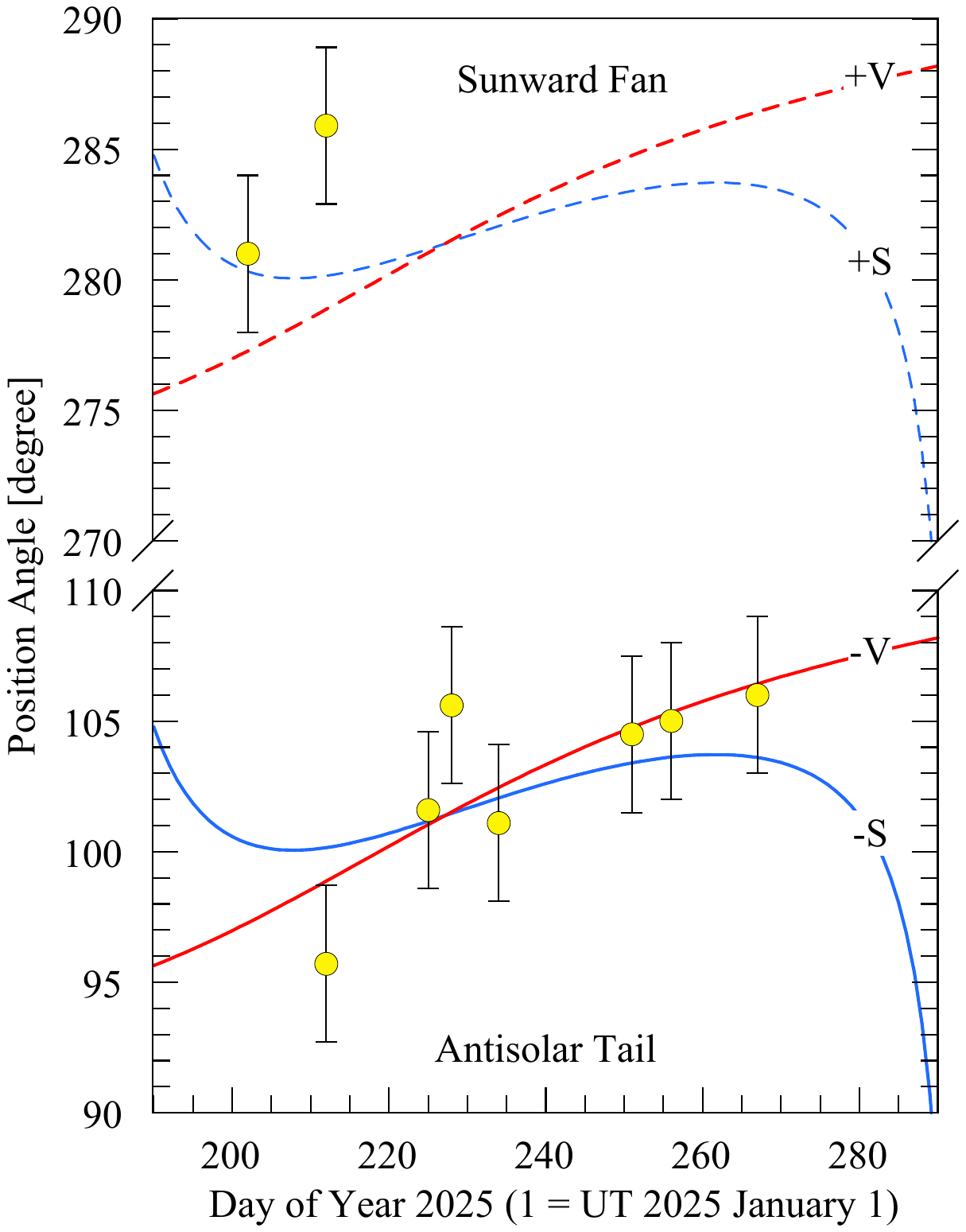}

\caption{Yellow-filled circles show measurements of the position angles of asymmetries as a function of date, the latter expressed as Day of Year in 2025.  In the lower part of the plot the solid red and blue lines  marked ``-V'' and ``-S'' show, respectively, the projected anti-velocity vector and antisolar directions.  The dashed red and blue lines in the upper part of the plot marked +V and +S are 180\degr~from their solid counterparts. Note that the vertical axis is broken between 110\degr~and 270\degr~for clarity of presentation.  \label{Angle_plot}}
\end{figure}

The measured position angles are shown in Figure \ref{Angle_plot} as a function of the date of observation.  Also shown in the figure are the projected antisolar and anti-velocity vectors (blue and red lines marked ``-S'' and ``-V'', respectively) from the Horizons ephemeris site\footnote{$\url{https://ssd.jpl.nasa.gov/horizons/}$}.  We also show directions 180\degr~opposite to -S and -V as dashed lines in the  upper part of the figure, marked ``+S'' and ``+V''.  In the earliest observations the coma extension is closest to the +S and +V directions  along position angles near 280\degr~to 285\degr.   In later observations  the tail aligns more closely with -S and -V, at position angles from 95\degr~to 106\degr.  Given the considerable uncertainties of measurement, we cannot determine whether the eastward (tail) extension is more closely aligned with the antisolar or negative velocity vectors, although there may be a slight preference for the latter.   This would be characteristic of large, slowly ejected particles which closely hug the orbit of the primary even as they recede from it. 

A very faint antisolar tail was measured in Hubble imaging from UT July 21 (DOY$_{25}$ = 202) but, at this time, the sunward fan still dominated the optical appearance of 3I.  In NOT data, the emergence of an obvious antisolar tail did not occur until about UT 2025 July 29 (DOY$_{25}$ = 210), fully a month after the first clear reports of activity on July 2 (DOY$_{25}$ = 183) \citep{Jew25b}.  We interpret this delayed appearance of the tail as evidence that the optically dominant dust particles in 3I are large, so experiencing only small  acceleration  from radiation pressure.  By the end of July, the tail is visible to $\sim$4\arcsec~from the nucleus, corresponding to 8$\times10^6$ m in the plane of the sky; with $\alpha = 13.8^o$ on 2025 July 31, the tail's length is $\ell = 8 \times 10^6 / \sin(\alpha) \sim 3\times10^7$ m if it is radial to the Sun.

Under constant anti-solar acceleration a particle ejected sunward at speed $u$ would reach the apex distance

\begin{equation}
X_R = u^2/(2\beta g_{\odot})
\label{XR}
\end{equation}

\noindent in time\begin{equation}
t_X = u/(\beta g_{\odot})
\label{tX}
\end{equation}

\noindent and thereafter be pushed down the tail. In these expressions,  $\beta$ is the size dependent radiation pressure efficiency factor and $g_{\odot}$ is the local solar gravitational acceleration. At the nominal heliocentric distance $r_H$ = 3 au we have $g_{\odot} = 6.7\times10^{-4}$ m s$^{-2}$.  For dielectric spheres, we assume $\beta \sim 1/a_{\mu}$ (c.f., \cite{Boh83}), where $a_{\mu}$ is the grain radius expressed in microns.  

\cite{Jew25} used a high resolution measurement of $X_R$ and Equation \ref{XR} to find 

\begin{equation}
u \sim 50 \beta^{1/2}.
\label{speed}
\end{equation}

\noindent Combining Equations \ref{tX} and \ref{speed}, the time from nucleus to apex is then

\begin{equation}
t_X \sim 7.5\times10^4/\beta^{1/2}~~\textrm{[s]}.
\label{tXb}
\end{equation}

To be pushed a distance $\ell$ down the tail relative to the nucleus in time $t$ implies a radiation pressure factor 

\begin{equation}
\beta = \frac{2(ut - \ell)}{g_{\odot} t^2}
\label{beta}
\end{equation}

\noindent in which distance $\ell$ is measured negative to the east of the nucleus.  Note that $t$ includes the time the particles take to travel to the apex of the sunward fan, and then turn around to be pushed antisunward into the tail.

   We combine Equations \ref{speed} and \ref{beta} to obtain a quadratic equation for $\beta$, into which we substituted $\ell = -3\times10^7$ m  (negative for distance down the tail) and $t \sim 2.6\times10^6$ s (1 month, the interval between the first emergence of barely detectable activity on July 1 and the unambiguous ground-based resolution of the tail by the end of the month).  We solve to find $\beta \sim$ 0.01 (implying $a_{\mu} \sim$ 100), giving $u$ = 5 m s$^{-1}$ by Equation \ref{speed}, while the  time to reach the apex is $t_X \sim$ 10 days, by Equation \ref{tXb}.   These estimates are corroborated by subsequent measurements of the emerging tail length.  For example, in our UT 2025 September 8 data, the tail is evident to at least 47\arcsec~from the nucleus, corresponding to $\ell = 2.0\times10^8$ if the tail is radial to the Sun.  This date is 70 days ($t = 6.0\times10^6$ s) from the presumed initial ejection on July 1.   Substitution again gives $\beta \sim$ 0.01 for the particles at the end of the tail on this date.  If particle ejection began substantially before July 1, then these estimates of $\beta$ should be regarded as upper limits and the inferred 100 $\mu$m particle size as a lower limit to the true value.  
   For example, the photometry by \cite{Mar25} shown in Figure \ref{Summary_plot} suggests that activity may have started as early as $\log_{10}(r_H)$ = 0.8 au (6.5 au, corresponding to May 1).  If so, $\beta$ would be smaller, and $a_{\mu}$ larger, by factors of 4 to 9.

   We conclude that the delayed emergence of the radiation pressure-swept tail in 3I is reasonably attributed to the relatively large size and low ejection velocity of the optically dominant particles, and to their correspondingly small acceleration by radiation pressure.  Similarly large particles were earlier inferred in 2I/Borisov \citep{Kim20} and are common in weakly active comets and active asteroids.  A possible explanation for the dominance of large particles is that smaller particles are more tightly bound by inter-particle forces and cannot be so readily detached  (\cite{Gun15}, \cite{Jew19}).

With $u$ = 5 m s$^{-1}$ the residence time for 100 $\mu$m dust particles in the 10$^4$ km radius photometry aperture employed in this analysis is $t_R \sim 2\times10^6$ s (23 days).  This long  timescale, during which the heliocentric distance of 3I decreases by 0.7 au to 0.8 au (Table \ref{geometry}), represents the averaging time for the photometry, such that short term fluctuations in the production rate from the nucleus (e.g., due to rotation) cannot be discerned.  Averaging in the dust coma may also contribute to the apparent invariance of the heliocentric index with distance, even as the CO$_2$/H$_2$O production ratio (measured as $\sim$8 at 3.3 au; \cite{Cor25}) is expected to decrease by an order of magnitude over the 4.4 $< r_H <$ 1.8 au distance range.

A weak water ice absorption was reported in the 3I coma at $r_H$ = 4.1 au by \cite{Yan25}. \cite{Ket25} assumed steady-state ejection and fitted a multiparameter sublimating grain model to the sunward fan using Hubble data from UT 2025 July 21 ($r_H$ = 3.8 au). They concluded that the grains are predominantly icy. Unfortunately, important results of this model (the effective  grain size, the average ejection speed, and the assumed albedo) are not given, inhibiting any comparison with the results presented here. In an update \cite{Ket25b} predict a maximum in the ice coma brightness near $r_H$ = 3.5 au for which the 3I data (Figure \ref{Photometry_Plot}) provide no evidence.

The mass of an optically thin collection of opaque spherical dust particles, $M_d$, is related to their total scattering cross-section, $C$, by  $M_d = 4/3\rho \overline{a} C$, where  $\overline{a}$ is the average particle radius and $\rho$ = 10$^3$ kg m$^{-3}$ is the assumed particle density. Cross-section $C$ is computed from the apparent magnitude measurements via the inverse square law, taking the red magnitude of the Sun as -27.09.  We adopt  $\overline{a}$ = 100 $\mu$m based on the current analysis and then obtain rough dust mass production rates from

\begin{equation}
\frac{dM_d}{dt} = \frac{4\rho \overline{a} C}{3 t_R}.
\label{dmbdt}
\end{equation}

\noindent For simplicity, we assume that the crossing time $t_R \sim 2\times10^6$ s is independent of $r_H$.  Derived values of $C$ and $dM_d/dt$ are listed in columns 5 and 6 of Table \ref{photometry}.  Listed uncertainties on $dM_d/dt$ in the Table account only for measurement errors on $m_R$ and neglect larger (but uncertain) systematic errors in the application of Equation \ref{dmbdt}. While clearly approximate and model-dependent (e.g., the 0.04 geometric albedo and the particle density needed to estimate dust cross-section and mass are both assumed, not measured), we note that the mass loss rate in Table \ref{photometry} is within a factor of 2-3 of $dM_d/dt$ = 120 kg s$^{-1}$ inferred for 100 $\mu$m particles in Hubble data taken UT 2025 July 21 ($r_H$ = 3.83 au, \cite{Jew25}) and with the 125 kg s$^{-1}$ rate of production of CO$_2$ on UT 2025 August 6 ($r_H$ = 3.33 au, \cite{Cor25}).  When extrapolated to the perihelion distance (1.356 au) the predicted dust production rate from 3I reaches 405$\pm$10 kg s$^{-1}$. For comparison, the second interloper, 2I/Borisov, reached perihelion at 2 au when it produced dust at about 70 kg s$^{-1}$ \citep{Kim20}.  3I dust production at 2 au is $\sim$180 kg s$^{-1}$ (Table \ref{photometry}).



\clearpage
\section{SUMMARY}

We used photometry within a fixed linear aperture to measure the pre-perihelion lightcurve of interstellar interloper 3I/ATLAS across the distance range 1.81 $\le r_H \le$ 4.45 au.

\begin{itemize}

\item The integrated light within a 10$^4$ km radius projected aperture, corrected for varying geocentric distance and phase angle, varies with heliocentric distance as a power law, $r_H^{-3.8\pm0.3}$, where the uncertainty is dominated by the unmeasured phase function.  This dependence is consistent with a dust production rate $\propto r_H^{-2}$, as expected from equilibrium sublimation of a supervolatile ice (and consistent with independent detection of a large CO$_2$ flux).  The heliocentric variation of 3I is distinct from that of the second known interstellar interloper 2I/Borisov, for which the integrated light follows $r_H^{-1.9\pm0.1}$ across a similar range of heliocentric distances, corresponding to a nearly constant coma cross-section.

\item Although different from each other, the heliocentric indices of 2I and 3I are not statistically distinct from the broad distributions of this index measured in both dynamically new and old solar system comets.  The index thus provides no constraint on the past thermal evolution of the interlopers.  

\item The dominant morphology of 3I/ATLAS changes from a sunward dust fan in observations taken before 2025 August to an antisunward tail on later dates.  The delayed emergence of the antisunward tail is attributed to the weak action of solar radiation pressure on large coma dust particles.  We infer an effective particle radius $\sim$100 $\mu$m and ejection velocity $\sim$5 m s$^{-1}$.  These properties are similar to 2I/Borisov, where large, slowly ejected dust grains dominated the optical appearance of the coma.

\item The dust mass loss rate in 3I/ATLAS is $\sim$180 kg s$^{-1}$ at 2 au, compared with $\sim$70 kg s$^{-1}$ in 2I/Borisov when measured at the same distance, and consistent with 3I having a larger or more active nucleus.  The extrapolated perihelion production rate from 3I is 405$\pm$10 kg s$^{-1}$.

\end{itemize}

\acknowledgments
We thank Anlaug Amanda Djupvik and the NOT observing team for making these observations possible.  Pedro Lacerda provided $n$ statistics from his comet sample.  We thank the anonymous referee for comments.





\begin{deluxetable}{llccrrrrrrcrrrr}
\tabletypesize{\scriptsize}
\tablecaption{Observations of 3I/ATLAS
\label{geometry}}
\tablewidth{0pt}
\tablehead{\colhead{DOY$_{25}$\tablenotemark{a}} & \colhead{Date\tablenotemark{b}} & \colhead{Time\tablenotemark{b}}  & \colhead{CAM\tablenotemark{c}} & \colhead{Exp\tablenotemark{d}}& \colhead{$r_H$\tablenotemark{e}}   & \colhead{$\Delta$\tablenotemark{f}} & \colhead{$\alpha$\tablenotemark{g}}  & \colhead{$\theta_{- \odot}$\tablenotemark{h}} & \colhead{$\theta_{-V}$\tablenotemark{i}} & \colhead{$\delta_{\oplus}$\tablenotemark{j}} & \colhead{$\nu$\tablenotemark{k}}     }

\startdata


183 & July 2 & 22:59 - 23:19&		STANCAM	& 12$\times$60 &	4.450&  3.448&  2.5&  114.7&  95.0&   -0.9& 281.1  \\	
184 & July 3 & 22:12 - 02:53  & STANCAM & 3$\times$60 & 4.415&  3.417&    2.9&   112.0&  95.1&   -0.9&  281.2 \\ 
185 & July 4  & 23:55 - 03:37 & STANCAM & 3$\times$60 &4.378&  3.385&    3.2&  109.7&  95.2&   -0.9&  281.4 \\
186 & July 5 & 23:42 - 03:10 & STANCAM & 3$\times$60 & 4.346&  3.357&   3.6&  108.1&  95.3&   -0.9&  281.6 \\
187 & July 6 & 23:40 - 02:52 & STANCAM & 7$\times$60 & 4.280&  3.302&  4.3&  105.6& 95.5&   -0.9&  281.9 \\
188 & July 7 & 23:35 - 01:40   & STANCAM & 5$\times$60 & 4.246& 3.275&   4.6& 104.7 &  95.6&    -0.9&  282.1 \\
192 & July 11 & 22:41 - 22:43    & ALFOSC & 2$\times$60 & 4.151&  3.201&   5.7&  102.7&  96.0&  -0.8&  282.6& \\
198 & July 17 & 22:38 - 22:39    & ALFOSC & 2$\times$60 & 3.953&  3.061&   8.0&  100.8& 96.8&  -0.7&  283.7 \\
202 & July 21 & 22:30 - 22:33    & ALFOSC & 4$\times$30 & 3.820&  2.979&    9.7&  100.2&  97.4&   -0.7&  284.5 \\
206 & July 25 & 22:44 - 22:50    & ALFOSC & 5$\times$30 & 3.689&  2.906&   11.3&  100.1&  98.0&   -0.6&  285.4\\
212 & July 31 & 22:42 - 22:45    & ALFOSC & 4$\times$30 & 3.493&  2.812&   13.8&  100.2&  99.0&   -0.4& 286.8 \\
219 & August 7 & 21:34 - 21:37   & ALFOSC & 4$\times$30 & 3.268&  2.727&  16.5& 100.7& 100.2&   -0.2&  288.7 \\
225 & August 13 & 21:52 - 21:56 & ALFOSC  & 4$\times$30 & 3.077&  2.670&  18.6&  101.3& 101.2&   -0.1&  290.5 \\
228 & August 16 & 21:09 - 21:13 & ALFOSC & 4$\times$30 & 2.983&  2.648&   19.6&  101.6& 101.7&    0.0&  291.5 \\
234 & August 22 & 21:38 - 21:41 & ALFOSC & 4$\times$30 & 2.793&   2.612&  21.2&   102.1&  102.6&    0.2&  293.7 \\
238 & August 26 & 20:32 - 20:35 & ALFOSC & 4$\times$30 & 2.671&  2.594 &  22.1&  102.5& 103.2&    0.4&  295.3 \\
241 & August 29 & 20:29 - 20:33 & ALFOSC & 4$\times$30 & 2.581&  2.583&  22.6&  102.8& 103.6&    0.5&  296.6 \\
251 & September 8 & 20:15 - 20:18 & ALFOSC & 4$\times$30 & 2.284&  2.555&    23.2&  103.4& 104.8&    0.8&  301.8 \\
256 & September 13 & 20:09 - 20:12 & ALFOSC & 3$\times$30 & 2.143&  2.545&   22.8&  103.6& 105.4&    0.9&  305.0 \\
265 & September 22 & 20:03 - 20:06 & ALFOSC & 3$\times$30 & 1.903&  2.525&   20.7&  103.6& 106.3&    1.2&  311.8 \\ 
267 & September 24 & 19:59 - 20:01 & ALFOSC & 3$\times$30 & 1.853&   2.519&   20.0&  103.6& 106.5&    1.2&  313.5 \\
269 & September 26 & 19:57 - 20:00 & ALFOSC & 3$\times$30 & 1.805&  2.514&   19.2& 103.4& 106.7&    1.3&  315.3 \\
273 & September 30 & 19:57 - 20:02 & ALFOSC & 18$\times$30 & 1.712&   2.501&    17.2&  103.0 & 107.0 &  1.4 & 319.3 \\
\enddata

\tablenotetext{a}{Day of Year, 1 = UT 2025 January 1}
\tablenotetext{b}{UT Date in 2025 and start times of the first and last observation, each night}
\tablenotetext{c}{Camera}
\tablenotetext{d}{Number of images $\times$ exposure time in seconds}
\tablenotetext{e}{Heliocentric distance, in au }
\tablenotetext{f}{Geocentric distance, in au }
\tablenotetext{g}{Phase angle, in degrees }
\tablenotetext{h}{Position angle of projected anti-solar direction, in degrees }
\tablenotetext{i}{Position angle of negative heliocentric velocity vector, in degrees}
\tablenotetext{j}{Angle of observatory from orbital plane, in degrees}
\tablenotetext{k}{True anomaly, in degrees}

\end{deluxetable}

\clearpage

\begin{deluxetable}{llccrcrrrcrrrr}
\tabletypesize{\scriptsize}
\tablecaption{Photometry
\label{photometry}}
\tablewidth{0pt}

\tablehead{\colhead{DOY$_{25}$\tablenotemark{a}} & \colhead{Date\tablenotemark{b}} & \colhead{$m_R$\tablenotemark{c}} & \colhead{$M_R$\tablenotemark{d}} & \colhead{$C$\tablenotemark{e}} & \colhead{$dM_d/dt$\tablenotemark{f}}  
}

\startdata
183 & July 2        & 17.55$\pm$0.05 & 14.76 & 635 & 42$\pm$2\\	
184 & July 3        & 17.59$\pm$0.05 & 14.81 & 601 & 40$\pm$2\\ 
186 & July 5        & 17.68$\pm$0.10 & 14.91 & 531 & 35$\pm$4\\
187 & July 6        & 17.60$\pm$0.08 & 14.83 &  550 & 37$\pm$3\\
188 & July 7        & 17.50$\pm$0.10 &  14.74 & 590 & 39$\pm$4\\
192 & July 11       & 17.16$\pm$0.02 & 14.41 &  767 & 51$\pm$1\\
202 & July 21       &  16.91$\pm$0.03 & 14.15 & 821 & 55$\pm$2\\
206 & July 25       & 16.69$\pm$0.02 & 13.92 & 947 & 63$\pm$1 \\
212 & July 31       &  16.51$\pm$0.04 & 13.71 & 1028 & 68$\pm$3\\
219 & August 7      &  16.25$\pm$0.03 & 13.41 & 1188 & 79$\pm$2\\
225 & August 13     &  16.08$\pm$0.03 & 13.20 & 1276 & 85$\pm$3\\
228 & August 16     & 16.01$\pm$0.02 &  13.11 & 1306 & 87$\pm$2\\
234 & August 22     & 15.83$\pm$0.02 & 12.90 &  1394 & 93$\pm$2 \\
238 & August 26     &  15.55$\pm$0.02 & 12.60 & 1682 & 112$\pm$2\\
241 & August 29     & 15.45$\pm$0.02 &  12.49 & 1739 & 116$\pm$2\\
251 & September 8   & 14.91$\pm$0.02 & 11.95 & 2240 & 149$\pm$3\\
256 & September 13  & 14.76$\pm$0.02 & 11.82 & 2214 & 148$\pm$3\\
265 & September 22  & 14.03$\pm$0.01 & 11.19 & 3116 & 208$\pm$2\\ 
267 & September 24  & 13.99$\pm$0.02 & 11.18 & 2963 & 197$\pm$4\\
269 & September 26  & 13.76$\pm$0.02 & 10.99 & 3372 & 224$\pm$4\\
273 & September 30  & 13.28$\pm$0.02 & 10.61 & 4340 & 289$\pm$6\\
\enddata


\tablenotetext{a}{Day of Year, 1 = UT 2025 January 1}
\tablenotetext{b}{UT Date}
\tablenotetext{c}{Apparent red magnitude and its uncertainty}
\tablenotetext{d}{Heliocentric magnitude, from Equation \ref{invsq2}}
\tablenotetext{e}{Scattering cross-section in km$^2$, albedo $p_R$ = 0.04 assumed}
\tablenotetext{f}{Dust mass loss rate in kg s$^{-1}$, from Equation \ref{dmbdt}}

\end{deluxetable}

\clearpage

\begin{deluxetable}{lllcrl}
\tabletypesize{\scriptsize}
\tablecaption{Model Fits to Pre-perihelion Photometry
\label{fits}}
\tablewidth{0pt}

\tablehead{\colhead{$r_H$\tablenotemark{a}} & \colhead{Facility\tablenotemark{b}} & \colhead{Photometry\tablenotemark{c}} & \colhead{$M_R$\tablenotemark{d}}   & \colhead{$n$\tablenotemark{e}} & \colhead{Reference\tablenotemark{f}}    }

\startdata
3I/ATLAS\\
4.6 to 1.8 & NOT & Fixed Radius (10$^4$ km) & 8.6$\pm$0.7 & 3.8$\pm$0.3 & This Work \\
7.7 to 2.6   &  ATLAS & Fixed Angle (10\arcsec)  & 8.1$\pm$0.7 & 4.2$\pm$0.4 & \cite{Ton25} \\
6.5 to 3.7 & ZTF & Fixed Angle (3\arcsec) & 9.3$\pm$0.3 & 3.9$\pm$0.1 & \cite{Ye25} \\
6.3 to 5.5 & TESS  &  Fixed Angle (63\arcsec)& -3$\pm$2 & 10.4$\pm$1.2 & \cite{Mar25} \\
~\\
2I/Borisov \\
2.6 to 2.0 & UH88  & Fixed Radius (10$^4$ km) & 12.4$\pm$0.1 & 2.0$\pm$0.1 & \cite{Hui20} \\
8.0 to 5.0 & ZTF & Fixed Angle (5\arcsec) & 12.3$\pm$0.5 & 1.9$\pm$0.2 & \cite{Ye20} \\
\enddata


\tablenotetext{a}{Heliocentric distance range of the photometry, au}
\tablenotetext{b}{Telescope}
\tablenotetext{c}{Measurement type: Fixed radius vs.~fixed angle}
\tablenotetext{d}{Heliocentric magnitude, c.f., Equation \ref{invsq2}}
\tablenotetext{e}{Heliocentric index,  c.f., Equation \ref{invsq2}}
\tablenotetext{f}{Reference for the measurements }

\end{deluxetable}


\clearpage

\clearpage

\clearpage 


\clearpage 


\clearpage 


\clearpage 


\clearpage 


\clearpage 


\clearpage 


\end{document}